# Introduction to Searching with Regular Expressions

Christopher M. Frenz

Department of Computer Engineering Technology

New York City College of Technology (CUNY)

300 Jay St

Brooklyn, NY 11201

Email: cfrenz@citytech.cuny.edu

## *Abstract:*

The explosive rate of information growth and availability often makes it increasingly difficult to locate information pertinent to your needs.  These problems are often compounded when keyword based search methodologies are not adequate for describing the information you seek.  In many instances, information such as Web site URLs, phone numbers, etc. can often be better identified through the use of a textual pattern than by keyword.  For example, many more phone numbers could be picked up by a search for the pattern (XXX) XXX-XXXX, where X could be any digit, than would be by a search for any specific phone number (i.e. the keyword approach).  Programming languages typically allow for the matching of textual patterns via the usage of regular expressions.  This tutorial will provide an introduction to the basics of programming regular expressions as well as provide an introduction to how regular expressions can be applied to data processing tasks such as information extraction and search refinement.

## *Introduction:*

The time period that we all live in is often described as the beginning of an information age, since the world's economic focus has started to shift away from the production of physical goods and instead is growing to revolve around the production and processing of information. Members of the computer industry have been a firsthand witness to the dawning of this information age, as they have helped to establish and maintain the vast array of computers, networks, and storage devices that make processing such large quantities of information possible. The benefits of this expansive proliferation of computer networks has been clearly apparent in that access to information is now often not more that a few key strokes or mouse clicks away. While almost no one will argue that the old adage "knowledge is power" holds true now more than ever, the ever increasing amounts of information being made available have led to new sets of challenges, with the main one being how does an individual or company separate out the information that is needed from the information that is not?

The scope of this challenge is readily apparent to anyone that has ever used a Web based search engine, in that even the most mundane queries likely return several thousand hits. In fact controlling access to information via search is considered so critical, that companies such as Google and Microsoft are all vying to take control of the search arena (Jones, 2006). Most Web based search methodologies are based on the idea of keywords in which the user types one or more words in the query box and the search engine tries to return all of the documents that contain the keywords specified. What generally differentiates these search engines is the quality of their ranking algorithms, which help to distinguish what documents are most relevant to the query submitter and what documents are of little relevance. Developing ways of improving such ranking algorithms is an area of active research interest (Al-Saffar & Heileman, 2007; Rungsawang et al., 2007), yet much of focus of these algorithms is still based upon the idea of improving keyword-based search methodologies.

Keyword based searches are highly useful for researching many topics, including information on the ranking of Web search results. For example, typing the query "page rank" into a search engine will return numerous hits that describe the PageRank algorithm used to weight search results returned by Google and some other search engines. Yet, what if the information you seek cannot readily be expressed in the form of a keyword, such as if you are trying to collect email addresses or phone numbers from a Web search engine (Note: this example is provided for illustrative purposes only and is not an endorsement of harvesting email addresses or phone numbers)? How could you express a phone number to Web search engine? If it's a specific phone number you seek it could easily be expressed by typing the actual number into the query box, but what if you don't know the number(s) and want to limit your search to web pages that only contain phone numbers? Perhaps you could try to add the keywords "phone number" to the query, but the rate of false positives and negatives is likely to be greater than if there was a direct way of expressing the format of a phone number in the query.

Human recognition of a phone number has to do with our ability to recognize the pattern of numbers that comprise a typical phone number. For example, within the U.S. all phone numbers follow some variant of the convention (XXX) XXX-XXXX, where X can be any digit. The key to the problem is to thus find a way for computers to be able to interpret and match textual patterns in the same way that they are able to match keywords.  Luckily, most modern programming languages already come equipped to this, by supporting regular expressions, which allow the development of text patterns for use in pattern matching.  For example, using the Perl 5 regular expression syntax (most modern regular expressions syntaxes are derivatives of this) a phone number could be matched with the expression:

```
/(\s?\(?\d{3}\)?[-\s.]?\d{3}[-.]\d{4})/
```

where the actual regular expression is contained between the set of /'s (Frenz, 2004; Frenz, 2005).

In this tutorial, the basic syntax and usage of regular expressions are going to be covered, as well a description of how regular expressions can be utilized to enhance search outcomes by programmatically processing Web search results through a regular expression based filter.

## *Regular Expressions*

Regular expressions are evaluated by computers using the concept of a state machine, or a process that will sequentially read in the symbols of an input word and decide whether the current state of the machine is one of acceptance or non-acceptance.  When you author a regular expression, you are defining all of the criteria that need to reach an accepted state in order for a valid pattern match to occur.  To see how this concept works lets define XYZ as our regular expression and try to perform a pattern match against the string ZXYXYZ.  At first, the symbol Z will be read into the state machine and will not match the first condition set forth by the state machine thus causing a state of non-acceptance and hence failure of the matching operation (Figure 1).

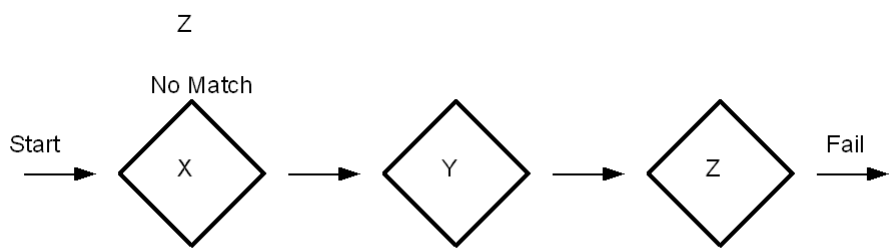

Figure 1: Z does not match the first condition specified by our XYZ state machine.

Since the Z symbol did not lead to an accepted state, the next character in the string (X) will be read into the state machine. Since X meets the first condition of the state machine, the next character (Y) is read in as well, which also meets the next condition of the state machine. Finally a third symbol (X) is read into the state machine which does not meet the final condition of the state machine and results in a failure to reach an accepted state (Figure 2).

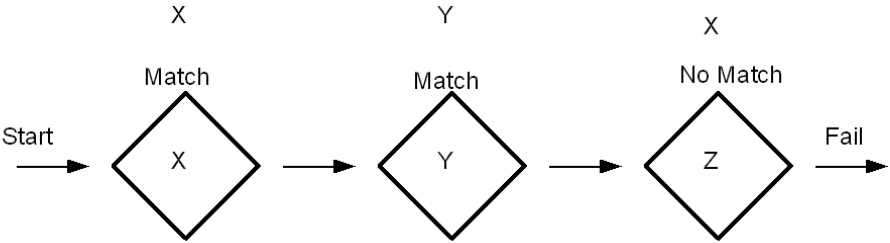

Figure 2: The symbols XYX do not match the conditions of the state machine.

Now that the portion of the string starting with second character failed to reach an accepted state, the third character of the string (Y) is used as a start symbol for the state machine. In this case the Y symbol fails to match the first condition of the start machine and results in a failure as well (Figure 3).

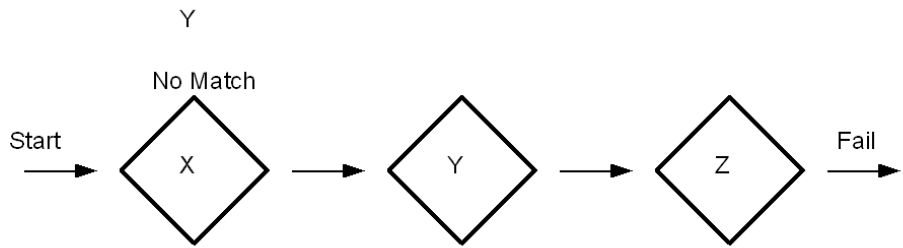

Figure 3: When Y is used as a start symbol, the string fails to match the state machine conditions

Finally, the next symbol in line (X) will be read into the state machine, which will successfully match the first condition. The remaining symbols Y and Z now meet the remaining conditions of the state machine and as such a condition of acceptance is reached by the state machine indicating a successful match of a string of characters to the regular expression (Figure 4).

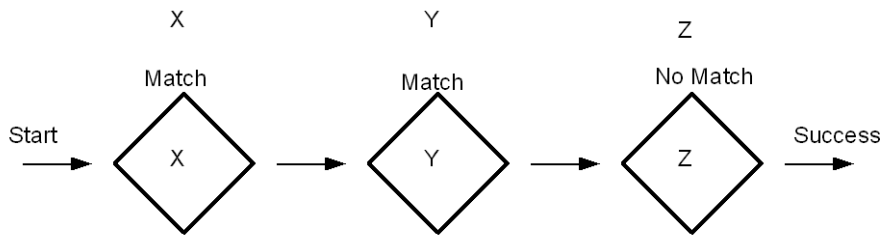

Figure 4: A successful pattern match.

More advanced state machines can be created by using the or operator (|) or parenthesis which allow for sub-patterns to be specified. For example, the regular expression XYZ|AB(C|c) would result in a state machine in which multiple branches could be used to produce an acceptance state (Figure 5).

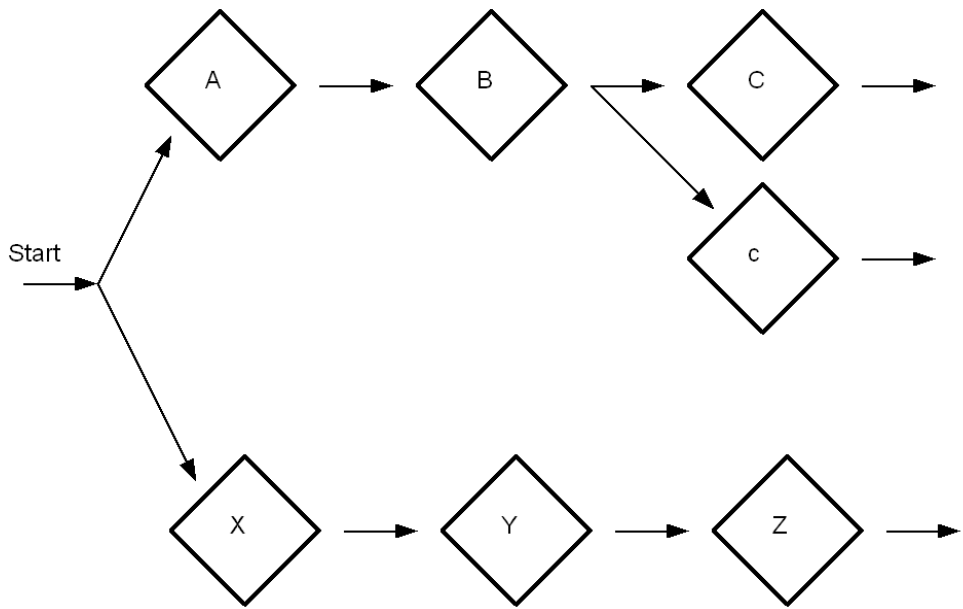

Figure 5: A state machine depicting a regular expression that contains a sub-pattern and or operators

This regular expression would thus allow the strings XYZ, ABC, or ABc to successfully match. The (C|c) portion of the expression is what allows for either an uppercase or lowercase C to be accepted following the letters AB. While the XYZ|AB(C|C) portion of the expression allows for the acceptance of either XYZ or AB(C|c) (Frenz, 2005; Freidl, 2006).

Regular expressions are typically utilized in conjunction with if statements or non-determinate loops as demonstrated in the following snippets of psuedocode.

if($mystring=~/mypattern/){

    # do some action

}

Or

```
while($mystring=~/mypattern/){

    #do some action

}
```

## Quantifiers

While it is possible to create almost any regular expression using individual symbols such as A, B, C, etc. there are often times that regular expressions can be shortened and simplified by using specialized regular expression operators such as quantifiers. Quantifiers allow you to match varying numbers of occurrences of a given symbol or sub-pattern. For example, If we wanted to match a pattern consisting of the character X one hundred times in a row, we could write out a regular expression consisting of 100 Xs or we could simply create a regular expression with a quantifier as follows X{100}. The various quantifiers available are listed in Table 1.

Table 1: Quantifiers that can be used in defining regular expressions.

| Quantifier | Effect |
| --- | --- |
| X* | Zero or more Xs |
| X+ | One or more Xs |
| X? | X is optional |
| X{3} | Three Xs |
| X{3,12} | Between 3 and 12 Xs |
| X{3,} | Three or more Xs |

When using quantifiers, one important thing to note is that by default Perl's regular expression engine is designed to be greedy in that it will always seek to find the biggest possible match so that if a match of the regular expression X[A-Z]*X was being performed against the string XABCXABCX, the regular expression would match the whole string and not just XABCX. This behavior can be changed by placing a ? after the quantifier, which will allow you to find the smallest possible match rather than the largest one. Thus if we sought to match XABCX the expression X[A-Z]*?X should instead be used.

One question that may arise is that the quantifiers could potentially be symbols that one is interested in matching as a part of a regular expression, and thus how could someone use a ? for instance as a part of a regular expression? By default Perl treats characters, such as ?, as metacharacters in that they have a special meaning to the regular expression engine. In order to turn off this behavior a metacharacter should be preceded by a backslash. Thus, adding \? to an expression would allow the ? to be considered part of the text pattern and not as a quantifier.

## Predefined Sub-patterns

In addition to quantifiers, there are other ways to simplify the time it takes to construct a regular expression and shorten the length of the expression. For example, if we were trying to match a phone number as previously mentioned, every time we wanted to match a digit of that phone number we could specify the pattern [0-9], or we could simply use the predefined sub-pattern \d. When combined with quantifiers these predefined sub-patterns become especially useful. For example, \d+ will match any positive integer or zero. The predefined sub-patterns commonly used in Perl regular expressions are demonstrated in Table 2, although there are other types of sub-patterns that can also be used such as POSIX character classes. Other programming languages that support regular expressions, such as the .NET languages, also support similar predefined subpatterns, although the syntax may vary (Frenz, 2004, Frenz, 2005, Freidl, 2006).

Table 2: Predefined Regular Expression Sub-patterns

| Sub-Pattern Specifier | Matching Pattern |
|---|---|
| \w | Any alphanumeric character or _ |
| \W | Any non-alphanumeric character or _ |
| \d | Any digit |
| \D | Any non-digit |
| \s | \n, \r, \t, \f, or " " (e.g. Whitespace characters) |
| \S | Anything not matched by \s |
| . | Anything but \n |

**Capturing Substrings**

The basics of how regular expressions work has now been defined, as well as various ways to ease the development of regular expressions via quantifiers and predefined sub-patterns, but there is another useful purpose regular expressions can be used for beyond simple pattern matching, and that purpose is substring capturing. In Perl parenthesis serve two purposes, they allow sub-patterns to be defined as has already been demonstrated, but they also allow the substrings that match the enclosed patterns to be captured for further use or processing within the programs code. In Perl the first set of parenthesis opened (from left to right) will store it's contents in the variable $1, the second in $2, the third in $3, and so on. Thus for example if we consider a modified version of the phone number regular expression that was defined earlier

```
(\s?(\(?\d{3}\)?)[-\s.](?\d{3}[-.]\d{4}))
```

the first set of parenthesis would assign the entire number to $1, the second set would assign the area code to $2, and the third set would assign the remainder of the phone number to $3. When processing text-based data, substring capturing is often a highly useful ability in that it can allow pertinent information to be extracted from Web pages and other data sources (Frenz, 2005).

## *Using Regular Expressions as a Search Tool*

Now that a description of the basic functionality and usages of regular expressions has been provided, the question remains as to how such regular expressions can be applied to information searches to help to overcome the limitations that keyword based search methodologies can present. The simplest way to approach this problem is not go out and build your own search engine for use with regular expressions, but rather to use an existing search engine or data repository as a back-end for your own search program. In other words, you conduct a broad search using an existing search engine that should encompass all of the information that you are interested in and use your custom coded application to perform regular expression based pattern matching on the search results. This type of approach has been successfully implemented for Pubmed, one of the largest repositories of information pertaining to the biomedical literature. The utility of this approach was demonstrated by having a regular expression based search refinement program uncover almost all of the different types of mutations that can result in deafness within humans (Frenz, 2007).

The standard nomenclature for deafness mutations is the letters DFN followed by another letter which is followed by an integer. As previously mentioned, this pattern is easily expressed with a regular expression, such as DFN[A-Z]\d+, but hard to represent in a single keyword. Thus, Pubmed's inherent search capabilities were used to find all of the biomedical literature associated with the words "human ear" and the search refinement program then used to extract only the returned abstracts that contained the pattern specified by the regular expression. The utility of this approach was demonstrated in that the program correctly identified the 117 abstracts that contained the text pattern out of the total of 61,371 abstracts returned by Pubmed. The regular expression based search refinement was thus able to accurately reduce the results to just less than 0.2% of the original search space (Frenz, 2007).

While the example given above was from the domain of bioinformatics, this approach to searching is readily suitable for use with other search engines as well. Many search engines offer API's or other interfaces that allow search results to be directly downloaded into applications for further processing, such as Yahoo! Search Web Services (http://developer.yahoo.com/search/) or the Google AJAX Search API (http://code.google.com/apis/ajaxsearch/). These interfaces thus allow search results to be downloaded to a custom application where they can be further processed by regular expression based pattern matching and as such help to further refine the search results presented to the application user in ways that are not easily implemented using keywords alone.

When performing such regular expression-based search refinement, however, there are several potential caveats that one should consider when creating keywords for querying the search engine and when designing the regular expressions to be used for refinement. One such caveat is that it is important to ensure that the keywords used are broad enough to return all documents that are potentially relevant to your information needs.  For example, in the case of the deafness mutation search, one type of mutation was actually not returned by the regular expression-based search methodology because the abstract in which it appeared did not contained the words "human ear".  This is more of a limitation of the underlying keyword based search methodology than the regular expression based processing, but still demonstrates the importance of choosing appropriate keywords.  The caveat that is more directly related to the regular expression based refinement itself is the potential for false positives and false negatives.  The regular expression developer needs to ideally design a regular expression that will maximize matching to appropriate text elements (e.g. reduce false negatives), while minimizing the number of matches to inappropriate text elements (false positives) (Frenz, 2007).  In order to ensure that this is the case, it is a good idea to test out regular expressions on a representative data sample prior to using the regular expression in a search of a large data set.

In all, regular expression based pattern matching is a highly useful tool for augmenting keyword based search methodologies in that it can be used to dramatically reduce the number of search results the user will be presented with.